# Energy Efficient Software Matching in Vehicular Fog

Rui Ma, Amal A. Alahmadi, Taisir E. H. El-Gorashi, and Jaafar M. H. Elmirghani
*School of Electronic & Electrical Engineering, University of Leeds, LS2 9JT, United Kingdom*
e-mail: {ml16r5m, elaaal, t.e.h.Elgorashi, j.m.h.elmirghani}@leeds.ac.uk

**ABSTRACT**
Along with the development of Internet of Things (IoT) and the rise of fog computing, more new joint technologies have been proposed. Vehicular Ad-hoc Networks (VANET) are one of the emergent technologies that come with a very promising role, where the spare processing capabilities of vehicles can be exploited. In this paper, we propose a fog architecture to provide services for end users based on a cluster of static vehicles in a parking lot referred to as a vehicular fog. The proposed vehicular fog architecture is connected to the central data center through an optical infrastructure. As the processing requests from users require specific software packages that may not be available in all vehicles, we study the software matching problem of task assignments in vehicular fog. The goal of this paper is to examine the effect of software packages variety in vehicles on the assignment decision and the overall power consumption. A mixed integer linear programming (MILP) model was utilized to optimize the power consumption of the overall architecture, considering different numbers of software packages in the vehicular fog. The results reveal a power saving of up to 27% when vehicles are uploaded with four or more different software packages out of a library of ten software packages in this example.
**Keywords**: vehicular fog; software matching, power consumption; Mixed Integer Linear Programming (MILP)

## 1. INTRODUCTION

With the significant development of information and communication technologies (ICT), smart cities have been recognized as a notable area where IoT can play a significant role [1]. In particular, vehicular Ad hoc Networks (VANET) show a great potential in smart cities where the large number of vehicles in the city can be equipped with embedded computing, communication, and storage units. These connected vehicles can form a cluster and work as a fog processing node [2], [3]. As Fog computing operates at the network edge, and with more data offloaded to the fog, it has become an energy-efficient platform compared to the central data centers [4]. Although centralized cloud computing is widely used in modern networks to provide large computing resources, power consumption and cost problems face the cloud in the core network. Such centralized clouds are not suitable for services with strict requirements on delay [5]. Fog computing overcomes the latency issue in cloud computing through the location of fog units at the edge of the network which shortens the distance between users and processing units to reduce latency. Meanwhile the flexible distribution of fog nodes at the network edge improves the performance of the network [6]. Due to these features, fog computing can provide a method of supplying services for a growing number of smart end devices. Due to the wide variety of end device applications and the limited resources in the vehicles as processing nodes, vehicles cannot install all the applications required by the large number of users in a smart city. Therefore, the software matching problem must be addressed and considered.

Recent research efforts have been focused on achieving energy-efficient service provisioning and fulfilling end-user application demands through conventional data centers [7]–[14] or different distributed fog data centers including static servers [4], [15], IoT nodes [16], [17], mobile devices [18], [19], or vehicles [20]. Using vehicles in parking lots as a clustered distributed datacenter has become a very popular research topic [21] – [24]. Although these distributed small servers provide their embedded resources as a service, these resources are limited in capability, including the supported software packages corresponding to the applications requests [8]. This limitation could affect the application offloading decision and therefore the overall power consumption of the network, cloud and fog, which has not been addressed to the best of our knowledge in the vehicular fog computing context.

In this paper, we propose vehicular fog (VFog) as an integrating framework for fog computing and vehicular networking where vehicles share and pool their spare processing capabilities to form fog nodes at the network edge. With the deployment of VFog, vehicles equipped with on-board units (OBUs) constitute vehicular fog sharing their underutilized computing, communication and storage resources which works as an infrastructure at the network edge. The remainder of this paper is organized as follows: In Section 2, the proposed vehicular fog framework is developed and the energy-efficient software matching problem is modeled. The model results are presented and discussed in Section 3. Finally, Section 4 concludes the paper and summarizes some of the future work directions.

## 2. SOFTWARE MATCHING PROBLEM IN VEHICULAR FOG FRAMEWORK

Figure 1 shows the proposed vehicular fog (VFog) architecture, where parked vehicles equipped with OBUs are clustered to form a fog computing node. The VFog is connected to the central cloud through an intermediate optical network which comprises of an IP over WDM core network and passive optical network (PON). The User Equipment (UE) can be any portable end device such as laptop, smart phone, wearable devices, smart watch and pad. UE plays the role of generating requests based on different task demands (required capacity and software type). The generated requests are wirelessly transmitted, to the nearby RSU which is located within the user range. The RSU is assumed to be aware of the available vehicles' resources and their different software packages. This RSU is equipped with a computing and communication unit to execute the offloading strategy and communicate with the potential vehicles in the VFog. After receiving requests from UE, the RSU locates the available vehicle resources according to the software-matching offloading strategy, and based on the resource allocation optimization model. The RSU then forwards requests to the access point, located in the parking lot, which can communicate with the available vehicles and forward requests to their OBU's processor. Each vehicle is preloaded with one or more software packages, to support different types of requests and to satisfy user requests. In case of VFog assignment failure, because of resource or software packages limitations, the RSU forwards the requests to the central cloud to fulfill tasks processing.

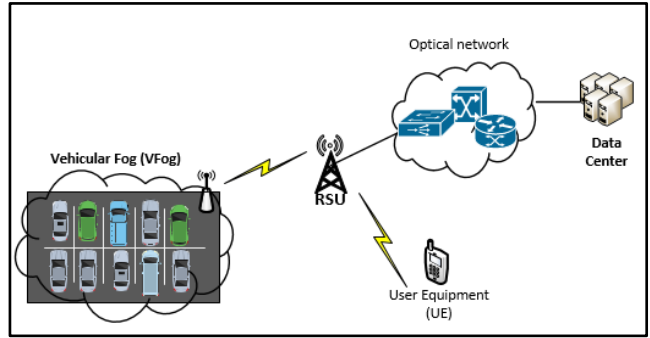

*Figure 1. Vehicular Fog Framework.*

The optimization model is developed, using mixed integer linear programming (MILP), with the goal of minimizing the power consumption considering the software matching problem. The model is examined through different scenarios where the vehicles could or could not satisfying all requests that demand different software types. The requests in this model are assumed to be assigned to only one processing unit (vehicle or central cloud).

In the proposed model, the objective is to minimize the total power consumed by both processing (by central cloud or vehicular fog) and networking (between RSU and central cloud/VFog), as defined by equation (1). Note that each request is defined by the triplet of (processing demand, data size, and software type).

$$\sum_{\substack{u \in U \\ i \in D}} CD_{ui} \cdot \delta_i + \sum_{\substack{u \in U \\ v \in V \\ s \in S}} PA_{uvs} \cdot \delta_v \quad (1)$$

where *U, V, S, D* are the sets of user requests, vehicles, software package types, and connected devices, respectively, $CD_{ui}$ is the data size demand for user request *u* transmitted through network device *i*. $\delta_i$ is the total energy per bit of the networking devices (intermediate devices, wireless devices, and vehicle), $PA_{uvs}$, is the processing demand assignment with application type *s* to vehicle *v* from user *u*. finally, $\delta_v$ is the energy per cycle of the vehicle processor.

The MILP model objective is subject to many constraints related to the task demands (processing and networking), software matching assignment, and the vehicular fog parameters (processing and link capacity). Due to paper length limitations, we omit the mathematical details of the constraints from this paper.

## 3. RESULTS AND DISCUSSION

In the proposed model, the number of user requests is assumed to be 50 requests. Processing demand values for each request are randomly assigned between 100 MHz and 300 MHz, and networking demands are considered to be proportional to the corresponding processing demand [25], [26]. All requests are processed, either by the central cloud servers, through the optical network, or by the vehicular fog processors, through wireless links whose capacity is equal to 450 Mbps [27]. The parking lot is assumed to have a maximum occupancy equal to 20 vehicles. All these vehicles participate by making their processing resources available. One type of embedded OBU is considered, and different number of software packages are considered. Different scenarios are considered with different number of software packages in each vehicle. Table I summarizes the main input values of the model.

*Table I. Input Data for the MILP Model*

| Device | Capacity | Power consumption |
|---|---|---|
| Vehicle OBU 1 | 240 MHz [28] | 3.6 W [28] |
| Access point | 800 Mbps [29] | 21.5 W [29] |
| RSU | 100 Mbps [30] | 15.5 W [30] |
| Central cloud server | 4000 MHz [8] | 300 W [20] |

Figure 2 shows the total power consumption of the network in different scenarios where the number of software packages in each vehicle ranged between zero and ten packages. When no software package is available (or specified) in any vehicle, all tasks are processed by the central cloud. Therefore, the power consumption reaches the highest value. That power consumption decreases as the numbers of different software packages available in each vehicle increases and achieves its minimum value when each vehicle contains 4 or more different software packages. This is because when 4 to 10 different software packages are available in each vehicle, the diversity of software packages in VFog is large enough to satisfy the optimal number of processed tasks that the MILP needs to base in the vehicular Fog. Compared to the case where all requests are processed by the central cloud, the maximum power saving reaches 27% when no less than 4 software packages are loaded in each vehicle in this example.

We also evaluate the change of workload in central cloud and VFog due to employing different software packages in vehicles, which is shown in Figure 3. Compared to the case where vehicles do not have software packages, the workload of the central cloud decreases significantly when the number of software packages is one in each vehicle. This is due to the availability of software packages in each vehicle which provides opportunities for local processing. Note that the number of vehicles is larger than the number of software types. Therefore, even one (different) software package per vehicle can quickly build a pool of different software packages in the vehicular car park. As a result, as the number of different software packages per vehicle increases, quickly the vehicular pool can acquire a full software library that contains all the software packages of interest. Accordingly, the workload of the cloud reaches its lowest value when each vehicle has 4 software packages installed. The workload values of VFog are the same when vehicles have more than 3 software packages. The decrease in the workload of the central cloud is 44% when VFog with 4 or more software packages per car is available and is used with the central cloud compared to the situation when all tasks are processed at the central cloud.

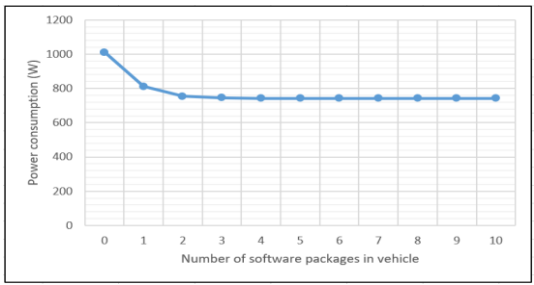

Figure 2. The total power consumption

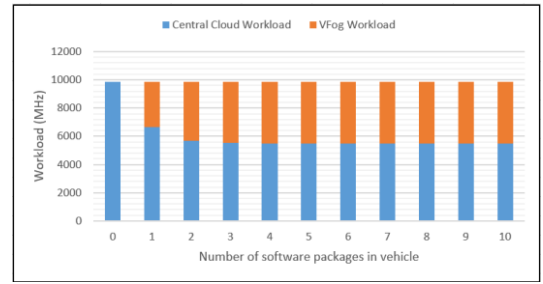

Figure 3. The workload of central cloud and VFog

## 4. CONCLUSIONS AND FUTURE WORK

In this paper we introduced a static vehicular fog framework considering the software matching problem in future intelligent transportation systems. The goal of this study is to minimize the power consumption through extending the processing to the network edge using the processing capabilities available in vehicles. We also investigate the software matching problem in VFog composed of vehicles in a parking lot. Optimizing the VFog through a MILP model, it was found that the total power consumption can be reduced by 27% when each vehicle has at least 4 software packages (out of a maximum of 10 software packages) installed compared to the situation when all tasks are processed by the central cloud. Meanwhile, the corresponding reduction in the central cloud workload is 44%. Future work includes considering different software packages popularities, different required processing power consumption for each software package, and the impact of processing delay for each loaded software.

**ACKNOWLEDGEMENTS**

The authors would like to acknowledge funding from the Engineering and Physical Sciences Research Council (EPSRC), through INTERNET (EP/H040536/1), STAR (EP/K016873/1) and TOWS (EP/S016570/1) projects. The second author would like to thank the Government of Saudi Arabia and Imam Abdulrahman Bin Faisal University for funding her PhD scholarship.